\documentclass{article}
\input{amssym}
\begin{document}
\title{{\bf Galilean Classification of Curves}}
\date{}
\author{Mehdi Nadjafikhah\thanks{Department of Mathematics, Iran University of Science and
Technology, Narmak-16, Tehran, Iran. e-mail:
m\_nadjafikhah@iust.ac.ir.}\and Ali Mahdipour Sh.\thanks{e-mail:
mahdi\_psh@mathdep.iust.ac.ir}}
\maketitle

\begin{abstract}
In this paper, we classify space--time curves up to Galilean group
of transformations with Cartan's method of equivalence. As an aim,
we elicit invariats from action of special Galilean group on
space--time curves, that are, in fact, conservation laws in
physics. We also state a necessary and sufficient condition for
equivalent Galilean motions.
\end{abstract}
\noindent A.M.S. 2000 Subject Classification: 53A55, 54H15, 83C40.
\section{Introduction}
Galilean transformation group has an important place in classic
and modern physics for instance: in quantum theory, gauge
transformations in electromagnetism, in mechanics \cite{AM}, and
conductivity tensors in fluid dynamics \cite{Ga}, also, in
mathematical fields such as Lagrangian mechanics, dynamics and
control theory \cite{Le}, and so on. In physics, when we study a
curve in Galilean space--time ${\Bbb R}^3\times{\Bbb R}$, it is
very important that we know about invariants of the curve, that
are conservation laws. In \cite{AM} for example, a Hamiltonian
vector field with some conditions, was introduced as a Galilean
invariant of special Galilean transformations on $T^{\ast}{\Bbb
R}^3$. But in this paper, by Cartan's method of equivalence
problem, we will find all invariants. We show that there are two
functionally independent invariants for a curve in a Galilean
space--time up to the action of special Galilean transformation
group, such that other invarians are functions of these invariants
and their derivations. Then, we use of this invariants to classify
space--time curves, in respect to special Galilean
transformations. In the next section, we state Cartan's theorem,
that is the main key for the classification. In section 3, we
propound the definition of Galilean group as a matrix group and
its properties. In the latest section, we determine the invariants
and classify space--time curves up to special Galilean group.
Finally, we prove that this invariants are a necessary and
sufficient condition for specification of space--time curves.
Then, we infer a physical result for Galilean motions, from this
classification.
%
%
\section{Preliminaries}
Let $G\subset{\rm GL}(n,{\Bbb R})$ be a matrix Lie group with Lie
algebra ${\goth g}$ and $P:G\rightarrow Mat(n\times n)$ be a
matrix-valued function which embeds $G$ into $Mat(n\times n)$ the
vector space of $n\times n$ matrices with real entries. Its
differential is $dP_B: T_BG\rightarrow T_{P(B)}Mat(n\times
n)\simeq Mat(n\times n)$.

\paragraph{Definition 2.1} The following form of $G$ is called {\it
Maurer-Cartan form}:
$$\omega_B=\{P(B)\}^{-1}\,.\,dP_B$$
that it is often written $\omega_B = P^{-1}\,.\,dP$.
The Maurer-Cartan form is the key to classifying maps into
homogeneous spaces of $G$, and this process need to this theorem
(for proof refer to \cite{Iv-La}):
\paragraph{Theorem 2.2 (Cartan)}{\it Let $G$ be a matrix Lie group with Lie
algebra ${\goth g}$ and Maurer-Cartan form $\omega$. Let $M$ be a
manifold on which there exists a ${\goth g}-$valued 1-form $\phi$
satisfying $d\,\phi = -\phi\wedge\phi$. Then for any point $x\in
M$ there exist a neighborhood $U$ of $x$ and a map $f:U\rightarrow
G$ such that $f^{\ast}\,\omega = \phi$. Moreover, any two such
maps $f_1, f_2$ must satisfy $f_1 = L_B\circ f_2$ for some fixed
$B\in G$ ($L_B$ is the left action of $B$ on $G$).}

\paragraph{Corollary 2.3 (\cite{Iv-La})}{\it Given maps $f_1, f_2:M\rightarrow
G$, then $f_1^{\ast}\,\omega =f_2^{\ast}\,\omega$, that is, this
pull-back is invariant, if and only if $f_1 = L_B\circ f_2$ for
some fixed $B\in G$.}\\

By corollary 2.3, one can conclude that in the view of Cartan's
theorem, the relation $f_1^{\ast}\,\omega =f_2^{\ast}\,\omega$
offers the invariant functions. In fact, these functions that we
call them invariants, when $f_1 = L_B\circ f_2$ for some fixed
$B\in G$, will remain permanent for maps $f_1$ and $f_2$ under the
pull-back action on Maurer-Cartan form $\omega$.

\section{Galilean Transformation Group}
Let ${\bf R}^3\times{\bf R}$ be a standard Galilean space--time. A
Galilean transformation is a transformation of ${\bf
R}^3\times{\bf R}$ as follows:
\paragraph{Definition 3.1} A map $\phi:{\bf R}^3\times{\bf R}\rightarrow{\bf R}^3\times{\bf
R}$ with the following definition
\begin{eqnarray*}
 \left(
\begin{array}{c} {\bf X}\\ t
\end{array}\right)\mapsto
\left(
\begin{array}{cc}
{\bf R}& {\bf v}\\ {\bf 0}^T & 1
\end{array}\right)\,\left(
\begin{array}{c} {\bf X}\\ t
\end{array}\right)+\left(
\begin{array}{c} {\bf y}\\ s
\end{array}\right)
\end{eqnarray*}
is called a {\it Galilean transformation}, where ${\bf R}\in{\rm
O}(3,{\Bbb R})$, $s\in{\Bbb R}$, and ${\bf y},{\bf v}\in{\Bbb
R}^3$.\\

The set of Galilean transformations is a 10-dimensional group
\cite{Iv-La}. We call this group as {\it Galilean transformation
group} or in brief, {\it Galilean group}, and denote it by ${\rm
Gal}(4,{\Bbb R})$. We can also identify this group with the
following matrix group
\begin{eqnarray}
{\rm Gal}(4,{\Bbb R})=\left\{ \left(\begin{array}{ccc} 1 & {\bf 0} & s \\ {\bf v} & {\bf R} & {\bf y} \\
0 & {\bf 0} & 1
\end{array}\right)\;\Big|\; {\bf R}\in{\rm
O}(3,{\Bbb R}),\; s\in{\Bbb R},\; \mbox{and}\;\; {\bf y},{\bf
v}\in{\Bbb R}^3 \;\right\},\label{eq1}
\end{eqnarray}
that with the matrix multiplication, is a 10-dimensional group.
Galilean group is a subgroup of affine transformation group
$A(5,{\Bbb R})$ and so a subgroup of ${\rm GL}(5,{\Bbb R})$. This
group also has a smooth structure and so is a smooth manifold.
Hence, with the smooth action of matrix multiplication, it is a
Lie group with Lie algebra ${\goth gal}(4,{\Bbb R})$. By its
representation, we can find its Maurer--Cartan forms that provide
a base for Lie algebra ${{\goth gal}}(4,{\Bbb R})$.
\paragraph{Definition 3.2} An element of Galilean transformation
group is called {\it special galilean transformation}, if in
representation (\ref{eq1}), ${\bf R}$ be in ${\rm SO}(3,{\Bbb
R})$. The group of all special Galilean transformations is called
{\it special Galilean transformation group} (or special Galilean
group in brief), and denoted by ${\rm SGal}(4,{\Bbb R})$. So, we
have
\begin{eqnarray*}
{\rm SGal}(4,{\Bbb R})=\left\{ \left(\begin{array}{ccc} 1 & {\bf 0} & s \\ {\bf v} & {\bf R} & {\bf y} \\
0 & {\bf 0} & 1
\end{array}\right)\;\Big|\; {\bf R}\in{\rm
SO}(3,{\Bbb R}),\; s\in{\Bbb R},\; \mbox{and}\;\; {\bf y},{\bf
v}\in{\Bbb R}^3 \;\right\}.
\end{eqnarray*}
${\rm SGal}(4,{\Bbb R})$ is a connected component of ${\rm
Gal}(4,{\Bbb R})$, and a Lie group with Lie algebra ${\goth s\goth
g\goth a\goth l}(4,{\Bbb R})$. In next section, we consider the
special Galilean group for classifying space--time curves, and
similar to Galilean group's, special Galilean group's
Maurer--Cartan form will be computed, it provides a base for Lie
algebra ${\goth s\goth g\goth a\goth l}(4,{\Bbb R})$.
\section{Classification of Space--time Curves}
 Let $c:[a,b]\rightarrow\Bbb{R}\times\Bbb{R}^3$ be a
curve with following definition:
$$c(t):=(t,{\bf X}(t))=(t,x_1(t),x_2(t),x_3(t)),$$
in which, the space coordinate ${\bf X}$, is a smooth
vector-valued function with values in $\Bbb{R}^3$, and $x_i$~s for
$i=1,2,3$, are a smooth scalar functions.\par
\paragraph{Definition 4.1} By {\sl ST--curve}, we mean a curve of class ${\cal C}^5$ that
 is in four dimensional space--time  ${\Bbb R}\times{\Bbb R}^3$,
 with this condition that it has no singular point, i.e. $\det({\bf X}',{\bf X}'',{\bf
 X}''')={\bf X}'\cdot({\bf X}''\times{\bf X}''')\neq0$. We may
 assume that this value be positive.\\

 If $c(t)=(t,{\bf X}(t))$ be a ST--curve, for all point $t\in[a,b]$
 we have ${\bf X}'(t)\neq 0$, and the curve ${\bf X}:t\mapsto{\bf X}(t)$ will be regular
 and one can reparameterize it with arc length parameter $s$, so that for each point $s$,
 we have $||{\bf X}'(s)||=1$.
 \paragraph{Definition 4.2} We call $c(t)=(t,{\bf X}(t))$ as {\sl regular}, if the curve ${\bf X}(t)$ be regular. Also,
 we say that the parameter of $c$ is {\sl arc length parameter}, if the parameter
 be an arc length parameter of ${\bf X}$.\\

The group of Galilean transformation can act on a ST--curve at
each point of the domain, when we equate $\Bbb{R}^3\times\Bbb{R}$
with
\begin{eqnarray*}
{\Bbb R}^5=\left\{\left(
\begin{array}{c}
t\\ {\bf X}\\1
\end{array}\right)\;\;\Big|\;\;t\in\Bbb{R},\,{\bf X}\in\Bbb{R}^3\right\},
\end{eqnarray*}
hence the action can be defined. Therefore, we say that {\sl two
ST--curves are equivalent if, their representations in ${\Bbb
R}^5$ be Galilean equivalent}.
\paragraph{Convention 4.3} Henceforth, we consider that image of
ST--curve $c$, be in ${\Bbb R}^5$ as above.  \par
We may consider a new curve $\alpha_c:[a,b]\rightarrow{\rm
Gal}(4,{\Bbb R})$ rather than $c$, in the following form:
\begin{eqnarray*}
\alpha_c(t)\!:=\!\left(\!\! \begin{array}{ccccc} 1 & 0 & 0 & 0 & t  \\[2mm]
{\bf X}' & \displaystyle{\frac{{\bf X}''}{||{\bf X}''||}} &
\displaystyle{\frac{{\bf X}''\times {\bf X}'''}{||{\bf X}''\times
{\bf X}'''||}} & \displaystyle{\frac{{\bf X}''\times({\bf
X}''\times {\bf X}''')}{||{\bf X}''\times({\bf X}''\times {\bf
X}''')||}} &  {\bf X}
\\[4mm]
0 & 0 & 0 & 0 & 1
\end{array} \!\!\right)
\end{eqnarray*}
where ${\bf X}$ is assumed as column matrix, and $||\;||$ is the
Euclidean norm. Obviously, for every time $t\in[a,b]$,
$\alpha_c(t)$ is an element of ${\rm Gal}(4,{\Bbb R})$ and so
$\alpha_c$ is well-defined.\par We can study $\alpha_c$ instead of
$c$, up to the action of Galilean transformation group as
following:\par
Let $c,\bar{c}:[a,b]\rightarrow{\Bbb R}^5$ be two ST--curves with
definitions $t\mapsto(t,{\bf X}(t),1)$ and
$\bar{t}\mapsto(\bar{t},\bar{{\bf X}}(\,\bar{t}\,),1)$
respectively. If $c$ be equivalent to $\bar{c}$ up to ${\rm
Gal}(4,(\Bbb R))$, that is, $\bar{c}=A\circ c$ for $A\in{\rm
Gal}(4,(\Bbb R))$, we have
\begin{eqnarray*}
\left(\begin{array}{c} \bar{t} \\ \bar{{\bf X}}\\ 1
\end{array}\right)= A \cdot\left(\begin{array}{c} t \\ {\bf X}\\ 1
\end{array}\right)=
\left(\begin{array}{ccc} 1 & {\bf 0} & s \\ {\bf v} & {\bf R} & {\bf y} \\
0 & {\bf 0} & 1
\end{array}\right)
\cdot\left(\begin{array}{c} t \\ {\bf X}\\ 1
\end{array}\right)
\end{eqnarray*}
then, we conclude that
\begin{eqnarray}
\bar{t}=t+s, \;\;\mbox{and}\;\; \bar{{\bf X}}={\bf R}\cdot{\bf
X}+t{\bf v}+{\bf y}\label{*}
\end{eqnarray}
First, second, and third differentiations of (\ref{*}) are in the
following form
\begin{eqnarray*}
\bar{{\bf X}}'&=&{\bf R}\cdot{\bf X}'+{\bf v}\\
\bar{{\bf X}}''&=&{\bf R}\cdot{\bf X}''\\
\bar{{\bf X}}'''&=&{\bf R}\cdot{\bf X}'''
\end{eqnarray*}
From above relations we have
\begin{eqnarray*}
\alpha_{\bar{c}}\! &=&
\!\left(\!\! \begin{array}{ccccc} 1 & 0 & 0 & 0 & \bar{t}  \\[2mm]
\bar{{\bf X}}' & \displaystyle{\frac{\bar{{\bf X}}''}{||\bar{{\bf
X}}''||}} & \displaystyle{\frac{\bar{{\bf X}}''\times\bar{{\bf
X}}'''}{||\bar{{\bf X}}''\times\bar{{\bf X}}'''||}} &
\displaystyle{\frac{\bar{{\bf X}}''\times(\bar{{\bf X}}''\times
\bar{{\bf X}}''')}{||\bar{{\bf X}}''\times(\bar{{\bf X}}''\times
\bar{{\bf X}}''')||}} & \bar{{\bf X}}
\\[4mm]
0 & 0 & 0 & 0 & 1
\end{array} \!\!\right)\hspace{0.8cm}\\
&=& {\tiny\hspace{-0.125cm}\left(\hspace{-0.25cm}
\begin{array}{ccccc}  1 & 0 &  0  & 0  &  t+s  \\[2mm]
{\bf R}\cdot{\bf X}' \!\!&\!\! \displaystyle{\frac{{\bf
R}\!\cdot\!{\bf X}''}{||{\bf R}\!\cdot\!{\bf X}''||}} \!\!&\!\!
\displaystyle{\frac{{\bf R}\!\cdot\!{\bf X}''\!\!\times\!\! {\bf
R}\!\cdot\!{\bf X}'''}{||{\bf R}\!\cdot\!{\bf X}''\!\!\times\!\!
{\bf R}\!\cdot\!{\bf X}'''||}} \!\! &\!\! \displaystyle{\frac{{\bf
R}\!\cdot\!{\bf X}''\!\!\times\!\!({\bf R}\!\cdot\!{\bf
X}''\!\!\times \!\!{\bf R}\!\cdot\!{\bf X}''')}{||{\bf
R}\!\cdot\!{\bf X}''\!\!\times\!\!({\bf R}\!\cdot\!{\bf
X}''\!\!\times\!\! {\bf R}\!\cdot\!{\bf X}''')||}} \!\! & \!\!
{\bf R}\!\cdot\!{\bf X}+ t\!\cdot\!{\bf v}+{\bf y}
\\[4mm]
0 & 0 & 0 & 0 & 1
\end{array} \hspace{-0.25cm}\right)}\\
&=^*& \left(\begin{array}{ccc} 1 & {\bf 0} & s \\ {\bf v} & {\bf R} & {\bf y} \\
0 & {\bf 0} & 1
\end{array}\right)\cdot\\
&&\hspace{0.8cm}\left(\!\! \begin{array}{ccccc} 1 & 0 & 0 & 0 & t  \\[2mm]
{\bf X}' & \displaystyle{\frac{{\bf X}''}{||{\bf X}''||}} &
\displaystyle{\frac{{\bf X}''\times {\bf X}'''}{||{\bf X}''\times
{\bf X}'''||}} & \displaystyle{\frac{{\bf X}''\times({\bf
X}''\times {\bf X}''')}{||{\bf X}''\times({\bf X}''\times {\bf
X}''')||}} &  {\bf X}
\\[4mm]
0 & 0 & 0 & 0 & 1
\end{array} \!\!\right)\\
&=& A\cdot\alpha_c,
\end{eqnarray*}
since the equation $=^*$ is concluded by knowing that for every
vectors ${\bf X}$ and ${\bf Y}$ in ${\Bbb R}^3$, and any element
${\bf R}\in{\rm SO}(3,{\Bbb R})$ we have ${\bf R}\cdot({\bf
X}\times {\bf Y})={\bf R}\cdot{\bf X}\times{\bf R}\cdot{\bf Y}$
and $||{\bf R}\cdot{\bf X}||=\det({\bf R})||{\bf X}||=||{\bf
X}||$. So we have
\paragraph{Theorem 4.4} {\it Two ST--curves $c,\bar{c}:[a,b]\rightarrow{\Bbb
R}^5$ are equivalent up to $A\in{\rm SGal}(4,\Bbb R)$ that is
$\bar{c}=A\circ c$; if and only if, the associated curves
$\alpha_c$ and $\alpha_{\bar{c}}$ are equivalent up to $A$ that is
$\alpha_{\bar{c}}=A\circ \alpha_c$.}\\

So our acclaim of working with $\alpha_c$ instead of $c$, does not
reduce the problem, with this benefit that we can use of Cartan's
theorem for $\alpha_c$ and then find its invariants. These
invariants in the view of theorem 4.4, are invariants of $c$.
Henceforth, we classify new curves $\alpha_c$~s up to ${\rm SGal}(4,\Bbb R)$.\\

From Cartan's theorem, a necessary and sufficient condition for
$\alpha_{\bar{C}}=B\circ\alpha_C=L_B\circ\alpha_C$ by $B\in {\rm
SGal}(4,{\Bbb R})$, is that for any left invariant 1-form
$\omega^i$ on ${\rm SGal}(4,{\Bbb R})$ we have
$\alpha_{\bar{C}}^{\ast}(\omega^i)=\alpha_{C}^{\ast}(\omega^i)$,
that is equivalent with
$\alpha_{\bar{C}}^{\ast}(\omega)=\alpha_{C}^{\ast}(\omega)$, for
natural ${\goth s\goth g\goth a\goth l}(4,{\Bbb R})$-valued 1-form
$\omega=P^{-1}\cdot dP$, where $P$ is the Maurer--Cartan form.\par
Thereby, we must compute the $\alpha_C^{\ast}(P^{-1}\cdot dP)$,
which is invariant under special Galilean transformations, that
is, its entries are invariant functions of ST--curves. This
$5\times 5$ matrix form, consists of arrays that are coefficients
of $dt$.\par
Since
$\alpha_C^{\ast}(P^{-1}\,.\,dP)=\alpha_C^{-1}\,.\,d\alpha_C$, so
for finding the invariants, it is sufficient that we calculate the
matrix $\alpha_C^{-1}\cdot d\alpha_C$. By differentiating of
determinant, we have
\begin{eqnarray*}
d\alpha_c(t)=
\left(\begin{array}{ccccc} 0 & 0 & 0 & 0 & 1 \\[2mm]
{\bf X}'' & A_1 & A_2 & A_3 &  {\bf X'}\\[4mm]
0 & 0 & 0 & 0 & 0
\end{array}\right)dt
\end{eqnarray*}
where we suppose that {\scriptsize\begin{eqnarray*} A_1
\!\!\!&=&\!\!\!\frac{{\bf X}'''||{\bf X}''||^2-{\bf X}''({\bf
X}''\!\cdot\!{\bf X}''')}{||{\bf X}''||^3},\\
A_2 \!\!\!&=&\!\!\! \frac{({\bf X}''\!\!\times\!\! {\bf
X}'''')||{\bf X}''\!\!\times\!\! {\bf X}'''||^2-\{({\bf
X}''\!\!\times\!\! {\bf X}''')\!\cdot\!({\bf X}''\!\!\times\!\!
{\bf X}'''')\}({\bf X}''\!\!\times\!\! {\bf X}''')}{||{\bf
X}''\!\!\times \!\!{\bf
X}'''||^3}, \\
A_3 \!\!\!&=&\! \!\!\frac{{\bf X}''\!\!\times\!\!({\bf
X}''\!\!\times\!\! {\bf X}'''')||{\bf X}''\!\!\times\!\!({\bf
X}''\!\!\times\!\! {\bf X}''')||^2\!}{||{\bf
X}''\!\!\times\!\!({\bf X}''\!\!\times\!\!
{\bf X}''')||^3}\\
&& -\frac{\!\{({\bf X}''\!\!\times\!\!({\bf X}''\!\!\times\!\!
{\bf X}''')\!\cdot\!({\bf X}''\!\!\times\!\!({\bf
X}''\!\!\times\!\! {\bf X}''''))\}({\bf X}''\!\!\times\!\!({\bf
X}''\!\!\times\!\! {\bf X}'''))}{||{\bf X}''\!\!\times\!\!({\bf
X}''\!\!\times\!\! {\bf X}''')||^3}.
\end{eqnarray*}}
We assumed that ${\bf X}$ is in the form $(x_1 \;\; x_2 \;\;
x_3)^T$. By knowing that $\det\alpha_c=1$, so the inverse matrix
of $\alpha_c$, $\alpha_c(t)^{-1}$, is in the form of following
matrix
{\tiny
\begin{eqnarray*}
\hspace{-0.35cm}\left(\!\!\!\!\!\! \begin{array}{ccc} 1 & {\bf 0}^T & -t  \\[2mm]
-\displaystyle{\frac{{\bf X}'\cdot{\bf X}''}{||{\bf X}''||}}
\!\!\!&\!\!\!\!\! \big\{\!\!\displaystyle{\frac{{\bf X}''}{||{\bf
X}''||}}\!\!\big\}^T
\!\!\!&\!\!\! \displaystyle{\frac{t({\bf X}'\!\cdot\!{\bf X}'')-{\bf X}\!\cdot\!{\bf X}''}{||{\bf X}''||}} \\[4mm]
-\displaystyle{\frac{{\bf X}'\!\cdot\!({\bf X}''\!\!\times\!\!{\bf
X}''')}{||{\bf X}''\!\!\times\!\!{\bf X}'''||}} &
\big\{\!\!{\displaystyle{\frac{{\bf X}''\!\!\times\!\!{\bf
X}'''}{||{\bf X}''\!\!\times\!\!{\bf X}'''||}}}\!\!\big\}^T \!\! &
\!\! \displaystyle{\frac{t({\bf X}'\!\cdot\!({\bf
X}''\!\!\times\!\!{\bf X}'''))-{\bf X}\!\cdot\!({\bf
X}''\!\!\times\!\!{\bf X}''')}
{||{\bf X}''\!\!\times\!\!{\bf X}'''||}}      \\[4mm]
\displaystyle{\frac{({\bf X}'\!\!\times\!\!{\bf X}'')\cdot({\bf
X}''\!\!\times\!\!{\bf X}''')}{||{\bf X}''||^2||{\bf
X}''\!\!\times\!\!{\bf X}'''||^2}A} \!\!&\!\!\!
\big\{\!\!\displaystyle{\frac{{\bf X}''\!\!\times\!\!({\bf
X}''\!\!\times\!\!{\bf X}''')}{||{\bf X}||^2||{\bf
X}''\!\!\times\!\!{\bf X}''')||^2}}\!\!\big\}^T \!\!\!&\!\!\!
-\displaystyle{\frac{t({\bf X}'\!\!\times\!\!{\bf
X}'')\!\cdot\!({\bf X}''\!\!\times\!\!{\bf X}''') -({\bf
X}\!\!\times\!\!{\bf X}'')\!\!\times\!\!({\bf
X}''\!\!\times\!\!{\bf X}''')}{||{\bf X}''||^2||{\bf X}''
\!\!\times\!\!{\bf X}'''||^2}A} \\[4mm]
0 & {\bf 0}^T & 1\end{array} \!\!\!\!\right),
\end{eqnarray*}}
where $A=||{\bf X}''\times({\bf X}''\times{\bf X}''')||$, and the
notation $^T$ means the transpose of a column matrix to be as a
row matrix.\par
After some straight computations, we find  $\alpha_c^{-1}\cdot
d\alpha_c$ as the multiplication of following matrix by $dt$:
\begin{eqnarray*}
\!\!\!\!\left(\!\!\!\!\!\begin{array}{ccccc}
0 & 0 & 0 & 0 & 1 \\[4mm]
||{\bf X}''|| & 0 & 0 & 0 & 0\\[4mm]
0 & 0 & 0 & \displaystyle{\frac{{\bf B}\cdot({\bf X}''\times{\bf
C})}{A\;||{\bf
X}''\times{\bf X}'''||}} &  0 \\[4mm]
\!\!\!\!\!\! 0 \!\!\!\!\!\!&\!\!\!\!\!
\displaystyle{\frac{A\;({\bf X}''\times{\bf B})\cdot{\bf X}'''}
{||{\bf X}''||^3||{\bf X}''\times{\bf X}'''||^2}}  &
\displaystyle{\frac{A\;({\bf X}''\times{\bf B})\cdot{\bf C}}
{||{\bf X}''||^2||{\bf X}''\times{\bf X}'''||^3}}
& 0 & 0 \\[4mm]
0 & 0 & 0 & 0 & 0
\end{array}\!\!\right)
\end{eqnarray*}
where ${\bf B}={\bf X}''\times{\bf X}'''$ and ${\bf C}={\bf
X}''\times{\bf X}''''$. We may assume that the parameter of
ST--curve be the natural parameter, the arc length $s$.
Henceforth, We adopt the problem with this assumption. Thus we
have $||{\bf X}'||=1$, then
\begin{eqnarray}
{\bf X}'\cdot{\bf X}''=0.\label{eq2}
\end{eqnarray}
Also if we assume that
\begin{eqnarray}
&&||{\bf X}''||=\mbox{constant},\label{eq3}\\
&&||{\bf X}'''||=\mbox{constant},\label{eq4}
\end{eqnarray}
By using of (\ref{eq2}) and (\ref{eq3}), we have ${\bf
X}'\cdot{\bf X}'''=-||{\bf X}''||=$ constant. By differentiating
of (\ref{eq3}) and (\ref{eq4}) in respect to $s$, we achieve that
${\bf X}''\cdot{\bf X}'''=0$ and ${\bf X}'''\cdot{\bf X}''''=0$,
respectively. Since $||{\bf X}''\times{\bf X}'''||^2=||{\bf
X}''||^2||{\bf X}'''||^2-({\bf X}''\cdot{\bf X}''')^2$, therefore
\begin{eqnarray}
&&||{\bf X}''\times{\bf X}'''||=\mbox{constant},\label{eq5}\\
&&{\bf X}''\cdot{\bf X}''''=-||{\bf
X}'''||=\mbox{constant}.\label{eq6}
\end{eqnarray}
From (\ref{eq3}) and (\ref{eq5}), we conclude that $A=||{\bf
X}''\times({\bf X}''\times{\bf X}''')||$ is also constant.\par

Since ${\bf X}''$ is perpendicular to ${\bf X}'''$, so ${\bf
X}''\times({\bf X}''\times{\bf X}''')=-{\bf X}'''$ and then we
find that
\begin{eqnarray}
({\bf X}''\times{\bf B})\cdot{\bf X}''' &=&-||{\bf X}'''||\nonumber \\
&=& \mbox{constant},\nonumber \\ ({\bf X}''\times{\bf B})\cdot{\bf
C} &=& {\bf B}\cdot({\bf X}''\times{\bf C})\nonumber \\
&=&{\bf X}''\cdot({\bf X}'''\times{\bf X}'''')\nonumber \\
&=&\sqrt{\det({\bf X}^{(i)}\cdot{\bf
X}^{(j)})}_{2\leq i,j\leq 4} \label{eq7} \\
&=&\!\!\sqrt{||\!{\bf X}'''||\,(||\!{\bf X}''||\,||\!{\bf
X}''''||\!+\!||\!{\bf X}'''||^2)}\nonumber \\
&=&\mbox{constant},\label{eq8}
\end{eqnarray}
where in the latest relation, the relation (\ref{eq7}) comes from
this fact that: for every vectors ${\bf X_1}$, ${\bf X_2}$, and
${\bf X_3}$ in ${\Bbb R}^3$, we have $\{{\bf X_1}\cdot({\bf
X_2}\times{\bf X_3})\}^2=\det({\bf X}_i\cdot{\bf X}_j)$. The
equation (\ref{eq8}) acquired from assuming that ${\bf
X}''\cdot({\bf X}'''\times{\bf X}'''')$ be positive (one can
consider the negative case), because the lengths $||{\bf X}''||$,
$||{\bf X}'''||$, and $||{\bf X}''''||$ do not vanish, therefore
${\bf X}''\cdot({\bf X}'''\times{\bf X}'''')$ is not zero.
Finally, we have
\begin{eqnarray*}
\alpha_c^{-1}(s)\cdot d\alpha_c(s)= \hspace{8cm}\\[2mm]
\!\!\!\!\!\!\!\left(\!\!\!\!\!\begin{array}{ccccc}
0 & 0 & 0 & 0 & 1 \\[4mm]
||{\bf X}''||& 0 & 0 & 0 & 0\\[4mm]
0 & 0 & 0 \!\!&\!\! \displaystyle{\frac{{\bf B}\cdot({\bf
X}''\times{\bf C})}{A\;||{\bf
X}''\times{\bf X}'''||}} &  0 \\[4mm]
\!\!\!\!\!\! 0 \!\!\!\!\!\!&\!\!\!\!\!
\displaystyle{\frac{-A\;||{\bf X}'''||} {||{\bf X}''||^3||{\bf
X}''\times{\bf X}'''||^2}}  \!\!&\!\! \displaystyle{\frac{A\;({\bf
X}''\times{\bf B})\cdot{\bf C}} {||{\bf X}''||^2||{\bf
X}''\times{\bf X}'''||^3}}
\!\!\!\!&\!\!\!\! 0 & 0 \\[4mm]
0 & 0 & 0 & 0 & 0
\end{array}\!\!\right)ds.
\end{eqnarray*}

Thereupon, by using of $s$, the arc length  as parameter, and
assuming that the second and third derivation of ${\bf X}$ (the
space coordinate of curve $c(t)=(t,{\bf X})$) have invariant
lengths, then the entries of the multiplications matrix in
$\alpha_c^{-1}\cdot d\alpha_c$ will be all invariant. By previous
description, although this invariants calculated for the curve
$\alpha_c$, but they are in fact invariants of the curve $c$.
We can summarize above results in following theorem:
\paragraph{Theorem 4.5} {\it Let $c:[a,b]\rightarrow\Bbb{R}\times\Bbb{R}^3$ be a ST--curve with
 definition $c(t):=(t,{\bf X}(t))$, then $\omega_1=||{\bf X}''||$ and $\omega_2=||{\bf X}'''||$ are
 differential invariants of $c$ up to special Galilean group ${\rm SGal}(4,{\Bbb R})$.
 In general, every other differential invariant of $c$, is
 functionally dependency to $\omega_1$, $\omega_2$, and their
 derivations in respect to the parameter.}
\paragraph{Remark 4.6} If we consider a ST--curve, the dimension of
its image is 1, but the dimension of $\Bbb{R}\times\Bbb{R}^3$ is
4, hence the number of invariants must be 3. In spite of finding
two fundamental invariants $\omega_1$ and $\omega_2$, one can add
another invariant for instance $\omega_3:=||{\bf X}''\times{\bf
X}'''||$, to complete set of essential invariants.
\paragraph{Theorem 4.7} {\it Let $c,
\overline{c}:[a,b]\rightarrow\Bbb{R}\times\Bbb{R}^3$ be two
ST--curves. $c$ and $\overline{c}$ are locally equivalent up to
special Galilean group ${\rm SGal}(4,{\Bbb R})$, if and only if,
$\omega_1=\overline{\omega}_1$ and
$\omega_2=\overline{\omega}_2$.}\\

\noindent{{\it Proof:}} Formerly, we proved that two curves which
are locally equivalent up to special Galilean transformation, have
same differential invariants. We prove the converse. Let $c$ and
$\overline{c}$ be two ST--curves on $[a,b]$ with representations
respectively $(t,{\bf X})$ and $(\overline{t},\overline{{\bf X}}
)$. We assume that $\omega_1=\overline{\omega}_1$ and
$\omega_2=\overline{\omega}_2$, we prove that there is a special
Galilean transformation $A\in{\rm SGal}(4,{\Bbb R})$, such that
$c$ and $\overline{c}$ in the view of convention 4.3, will be
special Galilean equivalent.\par
We suppose that images of $c$ and $\overline{c}$ be in ${\Bbb
R}^5$ by the convention. There is an element in ${\rm
SGal}(4,{\Bbb R})$ that transform one point of $c$ to one of
$\overline{c}$, because if we consider arbitrary points $(t_0,{\bf
X}_0,1)$ of $c$ and $(\overline{t}_0,\overline{{\bf X}}_0,1)$ of
$\overline{c}$, then there are unique ${\bf R}\in{\rm SO}(3,{\Bbb
R})$ and ${\bf y}\in{\Bbb R}^3$ such that $\overline{{\bf
X}}_0={\bf R}\cdot{\bf X}_0+{\bf y}$. Thus, there is the following
element of ${\rm SGal}(4,{\Bbb R})$ that transforms the first
point to the second:
\begin{eqnarray*}
\left(\begin{array}{ccc} 1 & {\bf 0} & \overline{t}_0-t_0 \\ {\bf 0} & {\bf R} & {\bf y} \\
0 & {\bf 0} & 1
\end{array}\right).
\end{eqnarray*}
So, we can assume that $c_1:=(t,{\bf X}_1,1)$ be a special
Galilean transformation of $c$, that intersects $\overline{c}$ in
the time $t_0\in[a,b]$, that is $c_1(t_0)=\overline{c}(t_0)$. Let
this special Galilean transformation be in the following form
\begin{eqnarray*}
\left(\begin{array}{ccc} 1 & {\bf 0} & s_1 \\ {\bf 0} & {\rm Id}_3 & {\bf y}_1 \\
0 & {\bf 0} & 1
\end{array}\right).
\end{eqnarray*}
Then, since there are unique $\widehat{{\bf R}}\in{\rm SO}(3,{\Bbb
R})$ and $\widehat{{\bf y}}\in{\Bbb R}^3$ so that transfer the
orthonormal frame
\begin{eqnarray*}
\left(\frac{{\bf X}_1''}{||{\bf X}_1''||}\,,\,\frac{{\bf
X}_1''\times{\bf X}_1'''}{||{\bf X}_1''\times{\bf
X}_1'''||}\,,\,\frac{{\bf X}_1''\times({\bf X}_1''\times{\bf
X}_1''')}{||{\bf X}_1''\times({\bf X}_1''\times{\bf X}_1''')||}
\right)(t_0)
\end{eqnarray*}
on $c_1(t_0)$ and the tangent vector ${\bf X}_1 '(t_0)$, to
orthonormal frame
\begin{eqnarray*}
\left(\frac{\overline{{\bf X}}''}{||\overline{{\bf
X}}''||}\,,\,\frac{\overline{{\bf X}}''\times\overline{{\bf
X}}'''}{||\overline{{\bf X}}''\times\overline{{\bf
X}}'''||}\,,\,\frac{\overline{{\bf X}}''\times(\overline{{\bf
X}}''\times\overline{{\bf X}}''')}{||\overline{{\bf
X}}''\times(\overline{{\bf X}}''\times\overline{{\bf X}}''')||}
\right)(t_0)
\end{eqnarray*}
on $\overline{c}(t_0)$ and $\overline{{\bf X}}_1'(t_0)$,
respectively. We suppose that $\widehat{c}:=(t,\widehat{{\bf
X}},1)$ be the curve introduced with the action of following
matrix of ${\rm SGal}(4,{\Bbb R})$ on $c_1$:
\begin{eqnarray*}
\left(\begin{array}{ccc}
1 & {\bf 0} & 0 \\
{\bf 0} & \widehat{{\bf R}} & \widehat{{\bf y}} \\
0 & {\bf 0} & 1
\end{array}\right).
\end{eqnarray*}
We suppose that the parameters of $\overline{c}$ and $\widehat{c}$
be arc length parameters (by the mean of definition 4.2). Now we
can replace the curves $\overline{c}$ and $\widehat{c}$ with their
corresponding curves $\alpha_{\overline{c}}$ and
$\alpha_{\widehat{c}}$ (respectively). So, if we prove that
$\alpha_{\overline{c}}=\alpha_{\widehat{c}}$, then we will have
$\overline{c}=\widehat{c}$. Moreover, we have
\begin{eqnarray*}
\alpha_{\widehat{c}}&=&\left(\begin{array}{ccc}
1 & {\bf 0} & 0 \\
{\bf 0} & \widehat{{\bf R}} & \widehat{{\bf y}} \\
0 & {\bf 0} & 1
\end{array}\right)
\left(\begin{array}{ccc}
1 & {\bf 0} & s_1 \\
{\bf 0} & {\rm Id}_3 & {\bf y}_1 \\
0 & {\bf 0} & 1
\end{array}\right)\alpha_c\\
&=&
\left(\begin{array}{ccc}
1 & {\bf 0} & s_1 \\
{\bf 0} & \widehat{{\bf R}} & \widehat{{\bf R}}\cdot{\bf y}_1+\widehat{{\bf y}} \\
0 & {\bf 0} & 1
\end{array}\right)\alpha_c,
\end{eqnarray*}
hence $\alpha_c$ and $\alpha_{\widehat{c}}$ are equivalent with an
element of ${\rm SGal}(4,{\Bbb R})$. Thereby, $\alpha_c$ and
$\alpha_{\overline{c}}$ will be equivalent, and by theorem 4.4,
the proof will be completed. Henceforth, we show that
$\alpha_{\overline{c}}=\alpha_{\widehat{c}}$. \par
For curves $\overline{c}$ and $\widehat{c}$ we have following
equations, respectively
\begin{eqnarray*}
d\,\alpha_{\overline{c}}&=&\alpha_{\overline{c}}\,\cdot\,\overline{b}\\
d\,\alpha_{\widehat{c}}&=&\alpha_{\widehat{c}}\,\cdot\,\widehat{b},
\end{eqnarray*}
when $\overline{b},\widehat{b}\in{\goth s\goth g\goth a\goth
l}(4,{\Bbb R})$. But for $c$ and $\widehat{c}$ from assumption, in
every point of domain we have $\omega_1=\overline{\omega}_1$ and
$\omega_2=\overline{\omega}_2$. Furthermore, in each point of
$[a,b]$.
\begin{eqnarray*}
\widehat{\omega}_1&:=&||\widehat{{\bf X}}''||=||\widehat{{\bf
R}}\cdot{\bf X}''||
=\det(\,\widehat{{\bf R}}\,)\,||{\bf X}''||=\omega_1\\
\widehat{\omega}_2&:=&||\widehat{{\bf X}}'''||=||\widehat{{\bf
R}}\cdot{\bf X}'''|| =\det(\,\widehat{{\bf R}}\,)\,||{\bf
X}'''||=\omega_2.
\end{eqnarray*}
So, we have $\widehat{\omega}_1=\overline{\omega}_1$ and
$\widehat{\omega}_2=\overline{\omega}_2$. Then, with above
expressions we conclude that for all points of $[a,b]$,
$\overline{b}$ and $\widehat{b}$ are same that we call it $b$.
Now, $\alpha_{\overline{c}}$ and $\alpha_{\overline{c}}$ are
satisfied in first order equations (resp.)
$d\,\alpha_{\overline{c}}=\alpha_{\overline{c}}\,\cdot\,b$ and
$d\,\alpha_{\widehat{c}}=\alpha_{\widehat{c}}\,\cdot\,b$, with the
initial condition
$\alpha_{\overline{c}}(t_0)=\alpha_{\widehat{c}}(t_0)$. Therefore,
we have $\alpha_{\overline{c}}(t)=\alpha_{\widehat{c}}(t)$ for all
$t \in[a,b]$, and so, proof is complete. \hfill $\diamondsuit$
\paragraph{Corollary 4.8} {\it In the physical sense, we can consider
that each ST--curve be a trace of a particle with mass $m$ and
under influence of a force ${\bf F}$. By theorem 4.7, we conclude
that:
\noindent{\bf 1.} Two particles with same masses
$m=\widetilde{m}$, under influence of forces ${\bf F}$ and
$\widetilde{{\bf F}}$ (resp.), have same trajectory, if and only
if, the norms of ${\bf F}$ and its derivative ${\bf F}'$ be equal
to the corresponding norms of $\widetilde{{\bf F}}$ and its
derivative $\widetilde{{\bf F}}'$.
\noindent{\bf 1.} In particular, we suppose that two observers
${\cal O}$ and $\widetilde{{\cal O}}$ move with accelerations
${\bf a}$ and $\widetilde{{\bf a}}$ (resp.), in an inertial
coordinate system. If we consider the paths of a particle (as
ST--curves) with mass $m$ in respect to observers ${\cal O}$ and
$\widetilde{{\cal O}}$, and under the effect of forces ${\bf F}$
and $\widetilde{{\bf F}}$ (resp.), then the pathes are equal under
a special Galilean transformation, if and only if, $||{\bf
F}||=||\widetilde{{\bf F}}||$ and $||{\bf F}'||=||\widetilde{{\bf
F}}'||$.}
{}


\begin{thebibliography}{}
\bibitem{AM} R. Abraham and J. E. Marsden, {\em Foundations of
Mechanics}, 2nd Edition, Addison-Wasley, Canada (1978).
\bibitem{Ga} G. Gallavotti, {\em Foundations of Fluid Dynamics},
Springer-Berlin, Heidelberg (2002).
\bibitem{Iv-La} T.A. Ivey and J.M. Landsberg, {\em Cartan for
Beginners: Differential Geometry via Moving Frames and Exterior
Differential System}, A.M.S. (2003).
\bibitem{Le} A.D. Lewis, {\em Lagrangian Mechanics, Dynamics, and Control},
Preprint available online at
\verb"http://penelope.mast.queensu.ca/~andrew/" (2003) .
\bibitem{O'n} B. O'Neill, {\em Elementary Differential Geometry},
Academic Press, London--New York (1966).
\bibitem{Sp} M. Spivak, {\em A Comprehensive Introduction to
Differential Geometry}, Vol. II and III, Publish or Perish,
Wilmington, Delaware (1979).
\end{thebibliography}
\end{document}